\RequirePackage{lineno} 
\documentclass[12pt]{iopart}  % for review and submission
\usepackage{graphicx}  % needed for figures
\usepackage{dcolumn}   % needed for some tables
\usepackage{bm}        % for math
\usepackage{amssymb}   % for math
\usepackage[ansinew]{inputenc}
\usepackage{hyperref}
\newcommand{\textsubscript}[1]{$_{\textrm{#1}}$}

\begin{document}

%\setpagewiselinenumbers
%\modulolinenumbers[5]
%\linenumbers

\title{Cyclic Depopulation of Edge States in a large Quantum Dot}

\author{S.~Baer \footnote{Author to whom any correspondence should be addressed}, C.~R\"ossler, T.~Ihn, K.~Ensslin, C.~Reichl, W.~Wegscheider} 
\address{Solid State Physics Laboratory, ETH Zurich, 8093 Zurich, Switzerland}
\ead{sbaer@phys.ethz.ch}

\date{\today}


\begin{abstract}
We investigate magneto-transport through a 1.6 $\mu$m wide quantum dot (QD) with adjacent charge detector, for different integer filling factors in the QD and constrictions. When this system is operated at a high transmission, it acts as a Fabry-Pérot interferometer, where transport is governed by a Coulomb blockade mechanism. For lower transmissions where the barriers are in the tunneling regime, we can directly measure the charge stability diagram of two capacitively and tunnel coupled Landau levels. The tunneling regime has been investigated in direct transport, as well as in single electron counting. The edge states within the dot are non-cyclically depopulated, which can be explained by a simple capacitive model and allows to draw conclusions about the edge state geometry within the quantum dot. 
\end{abstract}

%\pacs{}
\maketitle

%\section{\label{sec:level1}First-level heading}
% sections are not used for PRL papers

\section{Introduction}
Two-dimensional electron systems at low temperatures and in strong magnetic fields show a rich spectrum of highly degenerate, incompressible ground states \cite{stormer_fractional_1983,tsui_two-dimensional_1982}. Fractional quantum Hall states, occurring at a fractional filling factor $\nu$ with an odd denominator, are well described by the Laughlin wavefunction \cite{laughlin_anomalous_1983}. There exists a prominent exception from this hierarchy: the $\nu~=~\frac{5}{2}$ state \cite{willett_observation_1987}, which is believed to obey non-abelian statistics \cite{moore_nonabelions_1991,read_beyond_1999}. This remarkable property could make it an interesting candidate for the realization of a topological qubit \cite{nayak_non-abelian_2008}. Theoretical ideas for probing the statistics of the $\nu~=~\frac{5}{2}$ state are based on quantum dots, operated as Fabry-Pérot interferometers as a basic building block \cite{stern_proposed_2006,bonderson_detecting_2006,das_sarma_topologically_2005,ilan_coulomb_2008,bonderson_probing_2006}. 

Quantum dots exposed to a magnetic field also offer other interesting fields of study, as the investigation of the spin configuration \cite{rogge_multiple_2006} or few-electron addition spectra \cite{ciorga_addition_2000}. In the presence of a strong magnetic field, Coulomb blockade (CB) oscillations can no longer be described within a single-particle picture. Alternating compressible and incompressible regions are formed inside the dot, which can strongly modify the CB oscillations \cite{staring_periodic_1992}. Previous experiments have allowed the extraction of mutual capacitances of these regions for different filling factors \cite{heinzel_periodic_1994}. In those experiments, alternating high and low CB peak currents have been observed, which was attributed to a double dot-like behavior of two edge states inside the dot. However, for the interpretation of recent experiments using quantum dots as Fabry-Pérot interferometers \cite{zhang_distinct_2009,camino_quantum_2007,ofek_role_2010}, it is important to understand the detailed structure of edge states inside the QD and the parameter range, where this description is valid.\\
Here we present investigations of a large quantum dot with a quantum point contact (QPC) serving as a charge detector.  When the QD is operated as a Fabry-Pérot interferometer, we find resonances with a slope in voltage - magnetic field space and a periodicity characteristic for a Coulomb dominated effect, as already observed in previous experiments \cite{zhang_distinct_2009,camino_quantum_2007,ofek_role_2010}.
When the system is operated at a lower transmission where the barriers are in the tunneling regime, we observe a similar effect as in Ref. \cite{heinzel_periodic_1994}. However, the amplitude modulation can be observed over a large parameter range for different filling factors, allowing the direct measurement of the charge stability diagram of capacitively and tunnel coupled edge states.  As a consequence, we can estimate the width of the incompressible region separating the edge channels inside the QD. In contrast to previous experiments, this is accomplished by using capacitances, directly extracted from the measured charge stability diagram. Furthermore we are able to investigate the CB amplitude modulation by using (time-resolved) charge detection techniques, where it shows up as an increased / decreased tunneling rate. To our knowledge, single electron counting has never been performed with a QD of similar size. Direct transport measurements do not always reflect the full complexity of the edge state substructure inside a QD. In future experiments, single electron counting might provide additional important insight to charge localization and transport in micron-sized Fabry-Pérot interferometers.
Most proposed Fabry-Pérot interferometry experiments for probing properties of fractional quantum Hall states assume edge states to be one-dimensional electron or composite fermion channels with negligible interaction between compressible regions. We show, that when the edge states are confined to the QD, a complex behavior with compressible and incompressible regions is observed. The observed tunnel-coupling between the different compressible regions, i.e. the presence of tunnel-coupled alternative paths, might influence the outcomes of the proposed interferometry experiments.

\section{Experimental details}
The quantum dot has been fabricated on a Hall-bar, defined by wet-etching of a single-side doped GaAs/Al\textsubscript{\textit{x}}Ga\textsubscript{1-\textit{x}}As heterostructure with a mobility of $\mathrm{8.1\times 10^6~cm^2/Vs}$ and an electron density of $\mathrm{1.15\times 10^{11}~cm^{-2}}$. These structures employ a reduced proportion of Al in the spacer layer between the doping plane and the 2DEG (x=0.24 compared to x=0.30 or x=0.33 which are most widely used), which was shown to favor the formation of the $\nu~=~\frac{5}{2}$ state \cite{gamez_$nu5/2$_2011}. The electron gas resides 340 nm below the surface. Optical lithography, combined with chemical etching and metal evaporation are used to define the mesa, ohmic contacts and topgate leads. The quantum dot and charge detector gates are defined by electron-beam lithography with subsequent metal evaporation. Applying a negative voltage of $V_{\mathrm{G}}\approx\mathrm{-0.5 V}$ to the topgates depletes the electron gas underneath. Compared to double-side doped quantum wells with $\delta$-doped screening layers, this structure allows for a much better gateability \cite{rossler_transport_2011}: the conductance of a single QPC is non-hysteretic and stable in time.
Measurements have been conducted in a dry dilution refrigerator at a base temperature of $T_{\mathrm{MC}} \approx$ 10 mK and in magnetic fields between \textit{B} = 0 T and \textit{B} = 5 T, using standard four-terminal lock-in measurement techniques. A constant AC voltage modulation of an amplitude $<~20~\mu V$ has been applied via a current-to-voltage converter.

\section{Results and discussion}
\subsection{Zero magnetic field transport}
Fig. 1a shows the topgate layout of the quantum dot that has been used for the measurements. The two 800 nm wide QPCs with a channel length of 600 nm serve as tunnel barriers of the 1.6 $\mu$m wide quantum dot. The employed QPCs have shown to result in an almost harmonic confinement potential, apart from the regime very close to pinch-off \cite{rossler_transport_2011}. The special geometry ensures a smooth QPC potential which is believed to favor the self-consistent formation of separated edge states.  

%However, the increased channel length (most commonly channel lengths are of the range of $\approx$ 200 nm - 400 nm \cite{van_wees_quantized_1988,wharam_one-dimensional_1988}) makes it more prone to additional scattering centers inside. 

In addition to the plunger gate (PG) that allows for the tuning of the electrochemical potential of the QD, a QPC that serves as a charge detector (CD) \cite{field_measurements_1993,ihn_quantum_2009} has been implemented. When QPC$_1$ and QPC$_2$ are in the tunneling regime, finite-bias measurements give rise to characteristic Coulomb diamonds (Fig 1b), from which we extract charging energies of about 100 $\mu$eV. The Coulomb diamond measurement also demonstrates the good stability and control of the QD. Sharp kinks in the CD current $I_{\mathrm{CD}}$ (Fig. 1d), aligned with peaks in the dot conductance $G_{\mathrm{QD}}$ indicate addition / depletion of a single electron from the quantum dot. By pinching off the tunnel barriers even more, time-resolved single electron counting \cite{elzerman_single-shot_2004,schleser_time-resolved_2004,ihn_quantum_2009} is possible for rates below approx. 50 Hz.
\begin{figure}
\includegraphics[width=8cm]{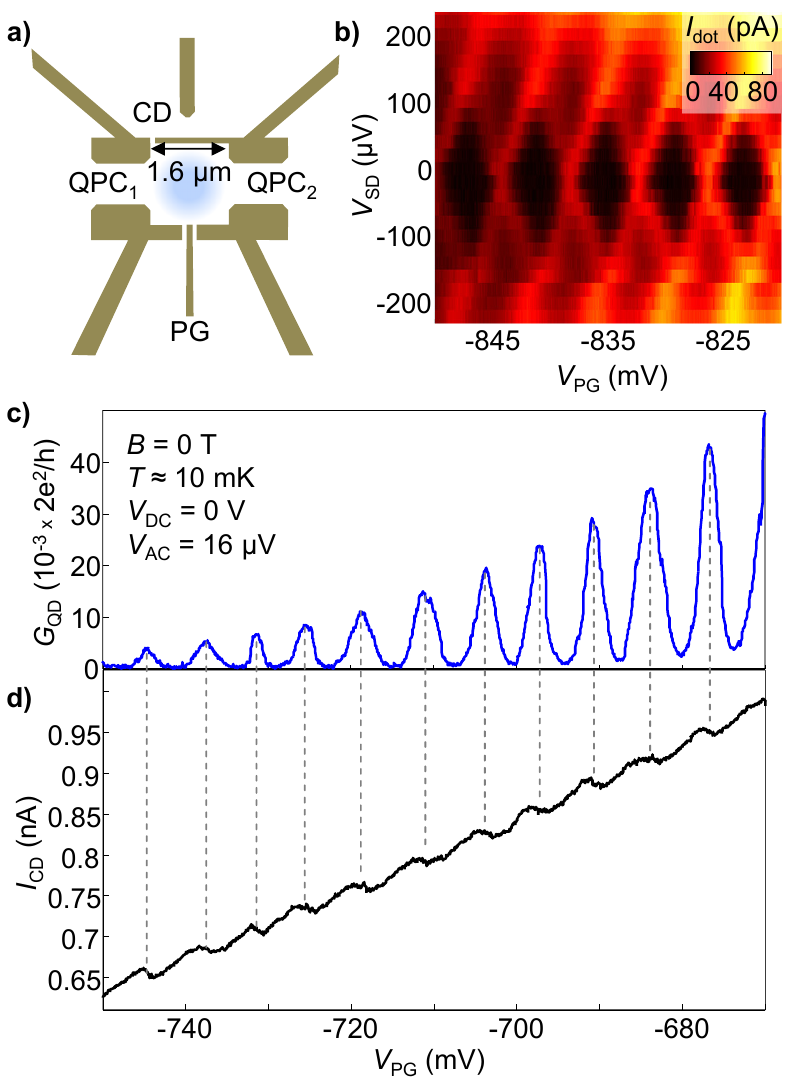}
\caption{\label{fig:1}\textbf{a):} Topgate layout of the quantum dot. The two 800 nm wide QPCs (QPC$_\mathrm{1}$ and QPC$_\mathrm{2}$) were designed to provide a smooth saddle-point potential. A third, 600 nm wide QPC serves as a charge detector (CD). The sample stability is sufficient for measuring regular and stable Coulomb blockade diamonds (\textbf{b}). Despite of the dot's large size, single electron charging events can be resolved in the charge detector (\textbf{d}) as well as in the direct current (\textbf{c}).}
\end{figure}

\subsection{Non-cyclic depopulation of edge channels} 
\begin{figure}
\includegraphics[width=8cm]{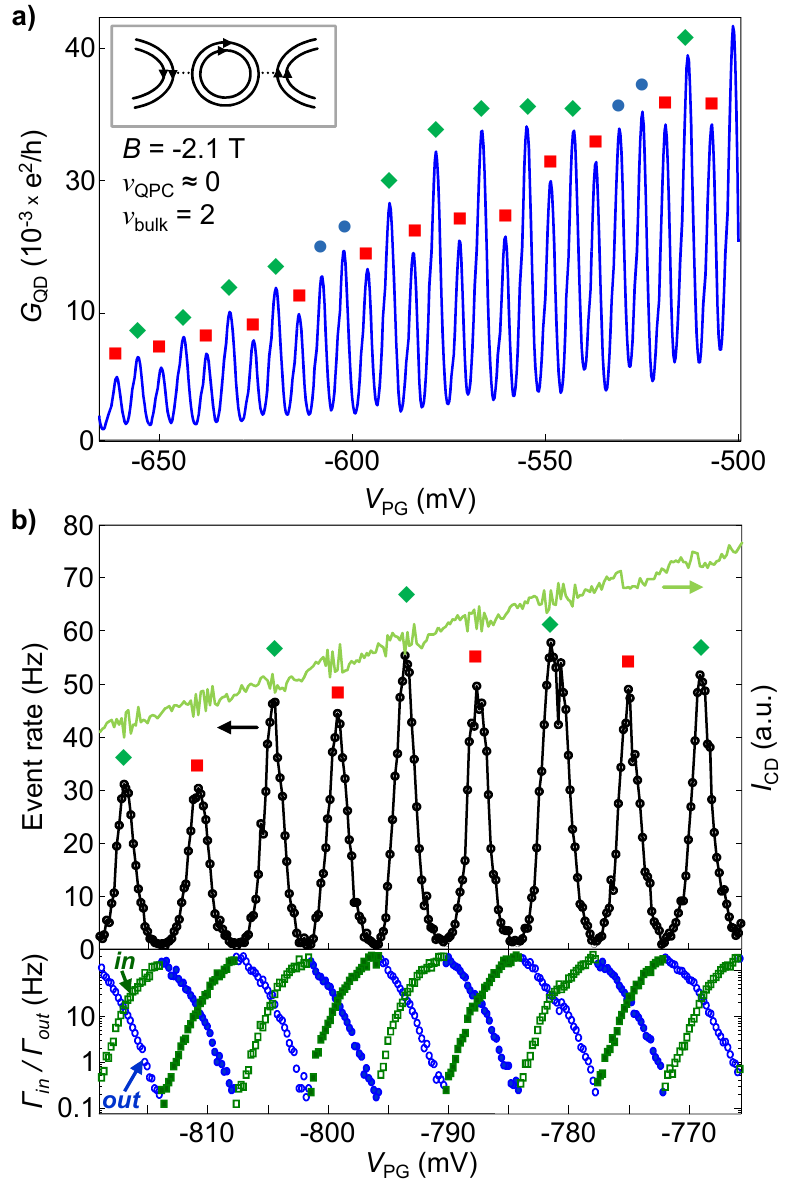}
\caption{\label{fig:2}\textbf{a):} Coulomb blockade peaks for a bulk filling factor of two. When depleting the dot with the plunger gate (PG), Coulomb peaks with a high / low peak current (diamonds / squares) are observed, interrupted by two peaks of similar magnitude (filled circles). The alternating pattern also shows up in the number of single electron charging events in a more pinched-off regime (\textbf{b}). Here, the rate of events is shown in black, with the corresponding charge detector current $I_{\mathrm{CD}}$ (shifted and scaled) in green (grey). Tunneling-in ($\mathit{\Gamma_{in}}$) and tunneling-out ($\mathit\Gamma_{out}$) rates for different Coulomb peaks are shown below.}
\end{figure}

For a bulk filling factor of $\nu_{\mathrm{bulk}} =2$, a filling factor in the quantum dot of $\nu_{\mathrm{QD}}\approx 2$ and $\nu_{\mathrm{QPC}}\approx 0$, i.e. when two spin-split edge states that are formed in the bulk, as well as in the dot are tunnel-coupled across the QPCs (see schematic inset Fig. 2a), CB oscillations are distinctly different from zero magnetic field measurements: the peak height of adjacent CB peaks alternates between two different values (Fig. 2a). After five peaks of high amplitude (marked by diamonds) and 6 peaks of low amplitude (squares), two peaks of similar height (filled circles) appear. The alternating peak height can also be seen in the single electron counting regime (Fig. 2b). Here, the rate of tunnelling events between dot, source and drain (black line, extracted from a time-resolved measurement of the charge detector current) is plotted as a function of the plunger gate voltage $V_{\mathrm{PG}}$. The contrast between peaks of high and low amplitude is lower than in the direct transport measurements. However, taking into account the amplitude dependence on the the plunger gate voltage (decreased Coulomb peak height as $V_{\mathrm{PG}}$ is decreased), we still have a peak height difference of roughly 15\%, bigger than our detection error. Tunneling-in ($\mathit{\Gamma_{in}}$) and tunneling-out ($\mathit\Gamma_{out}$) rates have been extracted from time-resolved measurements of the charge detector conductance. Apart from the different event rate at the Coulomb peak maxima, no further evidence of additional levels contributing to transport or more complex processes, like electron bunching \cite{gustavsson_counting_2006}, could be found. As argued later, this means that tunneling processes within the dot are much faster than processes between the quantum dot and the leads.
Measuring the CB oscillations as a function of PG voltage and magnetic field, a complex pattern of peak maxima is found (Fig. 3a). By extracting the peak amplitude minima (empty circles) and maxima (filled circles) numerically, it can be seen that they are distributed according to a tilted chessboard pattern, as indicated by the filled and empty circles in Fig. 3a. Along a Coulomb peak (black or blue (grey) line in Fig. 3a,b), the peak current is modulated as a function of the magnetic field (Fig. 3b). Neighboring Coulomb peaks show opposite amplitude dependencies. \\
Dominant shifts and drifts of the CB peaks in the measurement of Fig. 3a, make it impossible to investigate the absolute position of the peaks. However, modulations in the voltage spacing $\Delta V_{\mathrm{PG}}$ of two adjacent peaks can clearly be observed (Fig. 4c). Here, we plot $\Delta V_{\mathrm{PG}}$ of two successive CB peaks, offset in \textit{x}-direction for better visibility. Similar measurements can be conducted for a bulk filling factor $\nu_{\mathrm{bulk}}\approx 4$. In this case, bulk transport measurements suggest two spin-degenerate edge states separated by an incompressible region. Fig. 4d shows the voltage distance between adjacent CB peaks for this case.

\begin{figure}
\includegraphics[width=8cm]{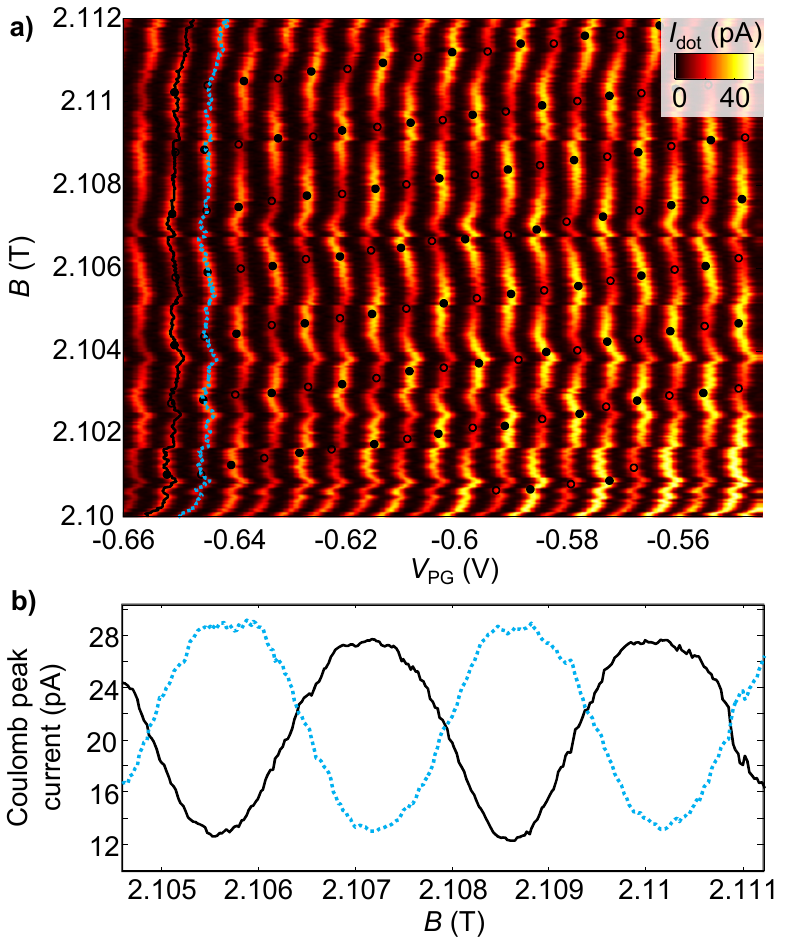}
\caption{\label{fig:3}\textbf{a):} Coulomb blockade peaks, measured as a function of magnetic field and voltage applied to the plunger gate. Minima and maxima of the Coulomb peak current are extracted numerically and indicated by filled / empty circles. \textbf{b):} Amplitude of two neighboring Coulomb peaks (position indicated in black / blue (grey) in \textbf{a}) as a function of the magnetic field. }
\end{figure}

The described behavior can be explained by a capacitive model \cite{heinzel_periodic_1994}: at a filling factor of $\nu_\mathrm{{QD}}~=~2$ in the quantum dot, the edge states corresponding to the spin-split lowest Landau level (LL) form compressible regions inside the quantum dot. We denote the lower / upper spin-split part of the lowest Landau level by LL1 and LL2. The width of these regions is dictated by self-consistency of the edge potential and the electrostatic potential contribution of the electron density \cite{chklovskii_electrostatics_1992}. In-between the compressible rings (shown schematically in Fig. 4a as thick lines), an incompressible region with a magnetic field dependent width is formed (hatched region). In this situation, electrons in both Landau levels populate the whole disk-shaped area in the quantum dot. However, only the compressible regions contribute to electron transport. Both spin-split levels inside the dot are tunnel coupled to source and drain and capacitively coupled to the plunger gate, as well as to the leads. In addition there is mutual capacitive coupling between the two spin-split levels. Although they overlap spatially, the electronic states are only tunnel-coupled via the compressible regions of both discs.
This configuration is an analog to a double quantum dot system, which in this case is formed by energetically separated, but spatially overlapping electronic states. Here, two main effects determine the configuration of the quantum dot: first, increasing the magnetic field increases the degeneracy of the Landau levels. For a constant number of electrons in the dot, this corresponds to a redistribution of electrons between LL1 and LL2. In addition to that, an increased magnetic field also leads to an increased spin splitting, translating to a larger separation and therefore reduced tunnel coupling in between the compressible regions. However, for the magnetic field ranges studied here, this tunnel coupling variation can be neglected. Second, the total population of the QD can be tuned via the plunger gate, which couples to both LL1 and LL2. Due to the spatial overlap and common center of mass of LL1 and LL2, we expect that the capacitive coupling of both regions to the plunger gate is similarly strong. The conversion factors between energy and gate voltage, the lever arms $\alpha_1$ and $\alpha_2$ for discs 1 and 2 thus are expected to be very similar, with a slightly bigger $\alpha_1$, considering the larger contribution to the capacitive coupling at the edge closer to the plunger gate. In this configuration, each spin-split level, LL1 and LL2, can be seen as a separate QD with single-particle energies $\frac{1}{2}\hbar \omega_c\pm\frac{1}{2}g\mu_BB$ and charging energies of $\frac{e^2}{C_1}$ and $\frac{e^2}{C_2}$, where $C_1$ and $C_2$ are the self-capacitances of discs 1 and 2. With the mutual capacitance $C_{1-2}$, the total energy of the double quantum dot with $N_1$ electrons in LL 1 and $N_2$ electrons in LL 2 can be expressed as:
\begin{eqnarray*}
E(N_1,N_2)&=&\frac{1}{2}\hbar\omega_c (N_1+N_2)-\frac{1}{2}g\mu_BB N_1+\frac{1}{2}g\mu_BB N_2 \\&&-e\alpha_1V_{\mathrm{PG}}N_1-e\alpha_2V_{\mathrm{PG}}N_2\\&&+\frac{e^2}{2C_1}N_1^2+\frac{e^2}{2C_2}N_2^2+\frac{e^2}{C_{1-2}}N_1N_2
\end{eqnarray*}
where $\omega_c=\frac{eB}{m^*}$ is the cyclotron frequency. The charge configuration of such a double quantum dot system can be described by a charge stability diagram with hexagonal regions of constant charge configuration ($N_1,N_2$) \cite{pothier_single-electron_1992,ruzin_stochastic_1992}. From the total energy, we may find conditions for the magnetic field and plunger gate voltage values along the boundary lines of this diagram (constant terms have been omitted):
\begin{tabular}{cl}\\
\hline transition& $B-V_{\mathrm{PG}}$ dependence\\
\hline$(N_1,N_2)\rightarrow (N_1+1,N_2-1)$&$B\varpropto\frac{e}{g\mu_B}\underbrace{(\alpha_2-\alpha_1)}_{\approx 0}V_{\mathrm{PG}}$\\
$(N_1,N_2)\rightarrow (N_1+1,N_2)$&$B\varpropto\frac{2e}{\frac{\hbar e}{m^*}-g\mu_B}\underbrace{\alpha_1}_{>0}V_{\mathrm{PG}}$\\
$(N_1,N_2)\rightarrow (N_1,N_2+1)$&$B\varpropto\frac{2e}{\frac{\hbar e}{m^*}+g\mu_B}\underbrace{\alpha_2}_{>0}V_{\mathrm{PG}}$\\\hline
\end{tabular}

\vspace{0.4cm} In Fig. 4b, such a charge stability diagram is shown schematically for given electron numbers in LL 1 and 2, $(N_1,N_2)$, as a function of the magnetic field $B$ and the plunger gate voltage $V_{\mathrm{PG}}$. Due to the comparable size of the capacitances $C_1$, $C_2$ and $C_{1-2}$, the hexagons have a nearly rectangular shape (from the measured charge stability diagrams explained later, it can be extracted that $C_{1-2}\approx0.87\times C_1$). Coulomb peaks occur, whenever charge configurations of $N_i$ and $N_i+1$ electrons on LL1 (\textit{i}=1) or LL2 (\textit{i}=2) are energetically degenerate. A high CB peak current is observed if the electrochemical potential of LL1 is aligned with the Fermi energy in source and drain, a low peak current corresponds to the alignment of the electrochemical potential of LL2 with the Fermi energy. The reason for this peak height modulation is the different lateral tunneling distance. 
The dashed (red) line in Fig. 4b indicates a $V_{\mathrm{PG}}$ trace, in which the amplitude difference between adjacent peaks is maximal (alternating transport via LL1 or LL2). Along this line, the charge degeneracy lines are crossed at a maximum distance from the triple points. In contrast, the dotted (green) line corresponds to a case, where charge configurations that contribute to high and low amplitude are energetically close. From the slope of these lines, we can conclude $\alpha_1<\alpha_2$, which might indicate a nonsymmetric charge distribution in the QD. Traversing the boundaries of the charge stability diagram near a triple point leads to thermal averaging of these two configurations, resulting in peaks of approximately equal height (as marked by the filled circles in Fig. 2a). Due to the slightly tilted hexagons, the high-low pattern is found again by further varying $V_{\mathrm{PG}}$. In addition, slightly different charging energies of the two LLs lead to a distorted hexagon pattern.

A change of magnetic field has two effects: for a situation with a constant total number of charges, the addition of flux quanta to the interior of the QD increases the degeneracy of both Landau level and their splitting, thus redistributing electrons between LL1 and LL2. A change of the magnetic field also influences the total population of the dot, as it shifts QD energy levels relative to the Fermi energy in the leads. The (red) dashed line in Fig 4c corresponds to the position of CB peaks with maximally modulated amplitudes. In agreement with the model illustrated in Fig. 4b, these lines correspond to approximately equal separation of adjacent peaks. For the second case, where the amplitude difference is thermally averaged (dotted (green) line in Fig. 4b), we expect and observe in Fig. 4c (along the green dotted line) alternating high and low $\Delta V_{\mathrm{PG}}$.

\begin{figure}
\includegraphics[width=8cm]{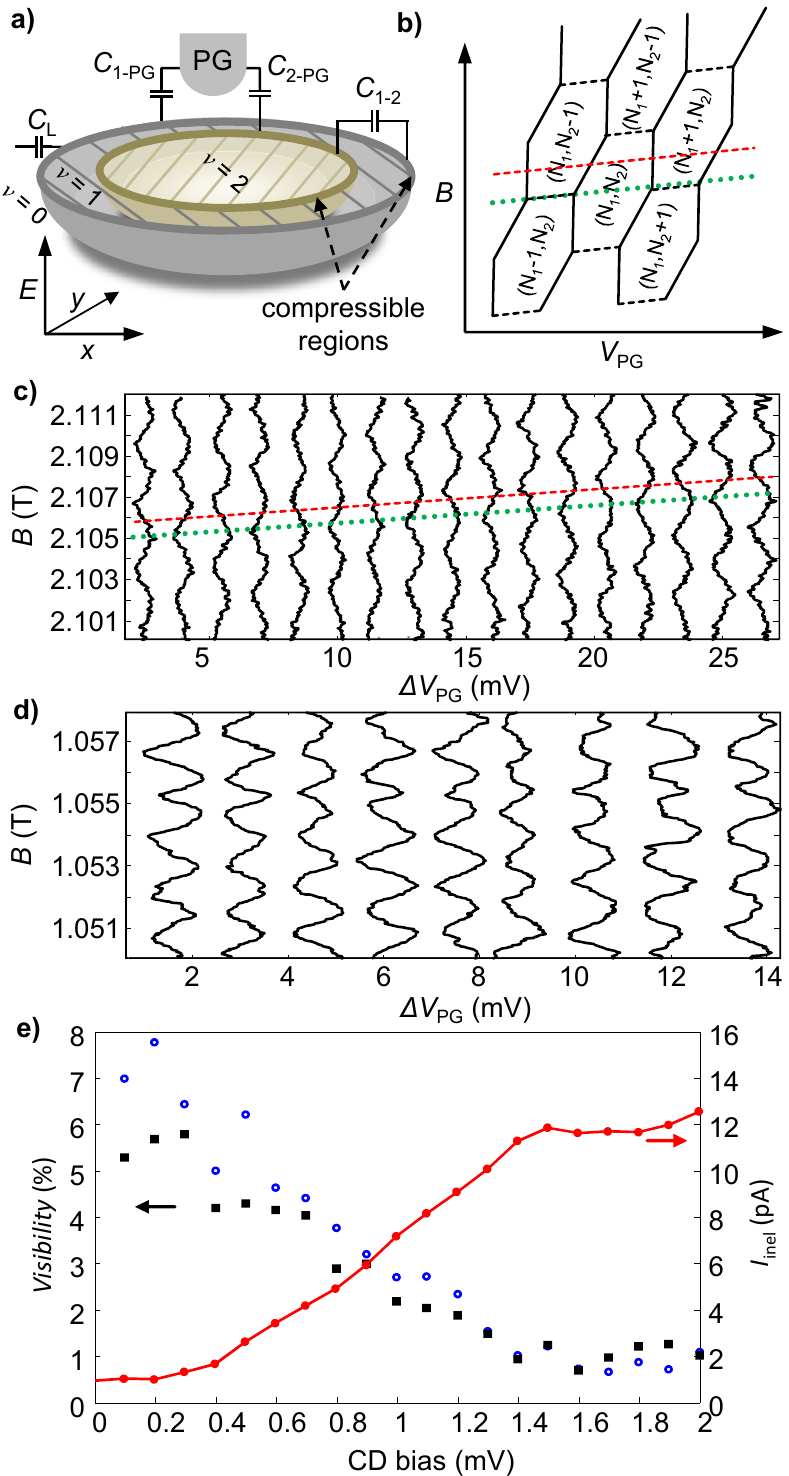}
\caption{\label{fig:4} \textbf{a):} Capacitive model for two coupled edge channels. The discoidal energy levels LL1 and LL2 are separated in energy, but overlap spatially. Tunneling of charges is possible in between the compressible regions where the Landau levels cross the Fermi energy. \textbf{b):} Exemplary charge stability diagram for a double quantum dot. Along the magnetic field axis, electrons mainly are redistributed in between LL1 and LL2 as well as slightly changing the total population by varying the total energy. A variation of the plunger gate voltage $V_{\mathrm{PG}}$ mainly influences the total electron population of the dot. The dashed (red) line indicates a situation in which the edge channels are cyclically depopulated, giving rise to a maximal height difference between neighboring peaks, as transport takes place alternatingly via LL1 or LL2. Along the dotted (green) line, neighboring Coulomb peaks lie close to the triple points and thus are thermally averaged and equally high. The plunger gate voltage difference between successive Coulomb peaks is shown in \textbf{c)} for $\nu_{\mathrm{QD}}\approx 2$ and \textbf{d)} for $\nu_{\mathrm{QD}}\approx 4$. The lines have been shifted closer together for better visibility. The dotted (green) line marks the position of Coulomb peaks of equal height, along the dashed (red) line, the peak height difference of neighboring peaks is maximal. \textbf{e):} Visibility of two Coulomb peak pairs as a function of the charge detector bias ($G_{\mathrm{CD}}\approx 0.25\frac{e^2}{h}$). Increasing the bias lowers the peak height difference visibility, while increasing the inelastic current through the quantum dot.}
\end{figure}

To distinguish if the amplitude modulation is caused by only different lateral tunneling distances, or an activated tunneling process, we can look at Fig. 4e:  here we measure the relative visibility ($(G_{LL1}-G_{LL2})/(G_{LL1}/2+G_{LL2}/2)$) of two thermally broadened pairs of Coulomb peaks, as a function of the bias that has been applied to the charge detector QPC at its maximum sensitivity (in our case $G\approx0.25\frac{e^2}{h}$ due to a localization in the QPC). The amplitude difference is observed to vanish when the bias is increased. The CD back-action is expected to increase the broadening of the Fermi-Dirac distribution of the leads. We are in the multilevel-transport regime ($h\Gamma\ll\Delta E\lesssim k_BT$, however not $\Delta E\ll k_BT$) \footnote{In this regime, the Coulomb peak conductance is expected to have a small temperature depencence (\cite{beenakker_theory_1991-1,kouwenhoven_electron_1997}), until either $k_BT\ll \Delta E$ or $k_BT\approx e^2/C$. (Note however, this does not hold whenever $h\Gamma\approx \Delta E$, where the amplitudes may have irregular and even nonmonotonic dependence on temperature \cite{meir_transport_1991}.)}.

The tunneling rate to both regions increases, as additional levels lie within the broadening of the Fermi-Dirac distribution. The broadening also leads to an increased occupation of the excited states of LL1 compared to the ground state of LL2 and thus an increased activated tunneling rate to LL2, which could explain the why the amplitude difference vanishes. From the 40\% maximum amplitude modulation between neighboring peaks, we can extract an energy-level separation of $\Delta E \approx 3~\mu$eV, using $\exp{(-\Delta E/k_{\mathrm{B}}T)}\approx 0.6$ and assuming a typical electron temperature of 60 mK. This is the order of magnitude expected for a dot of the given size.

One may ask, if there is any direct evidence that the second compressible region LL2 is involved in transport. In the situation where the electrochemical potential of LL2 is aligned with the potential of the leads (with a tunneling rate between LL1 and LL2 which is much slower than the tunneling rate between the leads and LL1), there are two sequential processes involved in an electron transfer from the leads to the QD: first, the fast activated tunneling of an electron to LL1 and back to the leads, second, slow tunneling from LL1 to LL2. Due to the very similar capacitive coupling of LL1 and LL2 to the charge detector, we are not able to resolve charge redistributions between those regions. From LL2, the electron can only escape with activated tunneling through LL1. While the electron has not left LL2, LL1 is blocked for further electron tunneling, due to the strong capacitive coupling $C_{1-2}$. The two tunneling processes would lead to electron bunching in the charge detector signal. However, such bunching is not observed in the experiment, suggesting that the interdot tunneling rate is very high (compared to the tunneling rate between QD and the leads) in our case.

Using the extracted interdot capacitance $C_{1-2}\approx 0.87~C_{1}$, we can make a rough statement about the spatial extent of the QD wave function. Modeling the interdot-capacitance as a simple plate-capacitor with a capacitance proportional to the plate area, we expect the area of LL2 to be roughly 87\% of the area of LL1.
The area of LL1 can be estimated from the lithographic size and the gate depletion lengths of the quantum dot, yielding $A\approx 0.64$ ($\mathrm{\mu}$m)$^2$. For rectangular QDs, this results in a difference of the side lengths of 54 nm. When the finite width of the edge states is neglected, this corresponds to a width of 27 nm of the incompressible region. Numerical calculations of bulk samples have predicted a width of approximately 20 nm for the incompressible region corresponding to a local filling factor of two \cite{lier_self-consistent_1994}. The enhanced value for our case could be a result of the simplicity of the model used which just allows for an order of magnitude estimate, or a smoother confinement potential and increased electron-electron interaction due to confinement. Similarly, a width of approximately 50 nm can be extracted from $C_{1-2}\approx 0.77~C_1$ in the case of $\nu_{\mathrm{QD}}\approx 4$. The increased width in this case is expected, as LL1 and LL2 are split by the larger cyclotron energy.

\section{Conclusion}
In summary, we have investigated transport through a large quantum dot, fabricated on a high-mobility wafer. Single-electron counting techniques, as well as direct current transport have been used to better understand the inner structure of the quantum dot for different filling factors. The periodic modulation of the conductance peak amplitude and spacing can be explained by a capacitive model, involving compressible and incompressible regions inside the dot. The high tunability of the device allowed the investigation of transport in the tunneling regime, as well as in a regime with edge states, perfectly transmitted through the dot (see appendix A). In this case, conductance oscillations, governed by a Coulomb blockade mechanism have been observed.

\section{Acknowledgments}
We acknowledge the support of the ETH FIRST laboratory and financial support of the
Swiss Science Foundation (Schweizerischer Nationalfonds, NCCR 'Quantum Science and Technology').
\clearpage

\appendix

\section{Transport in Fabry-Pérot regime}
\begin{figure}
\includegraphics[width=8cm]{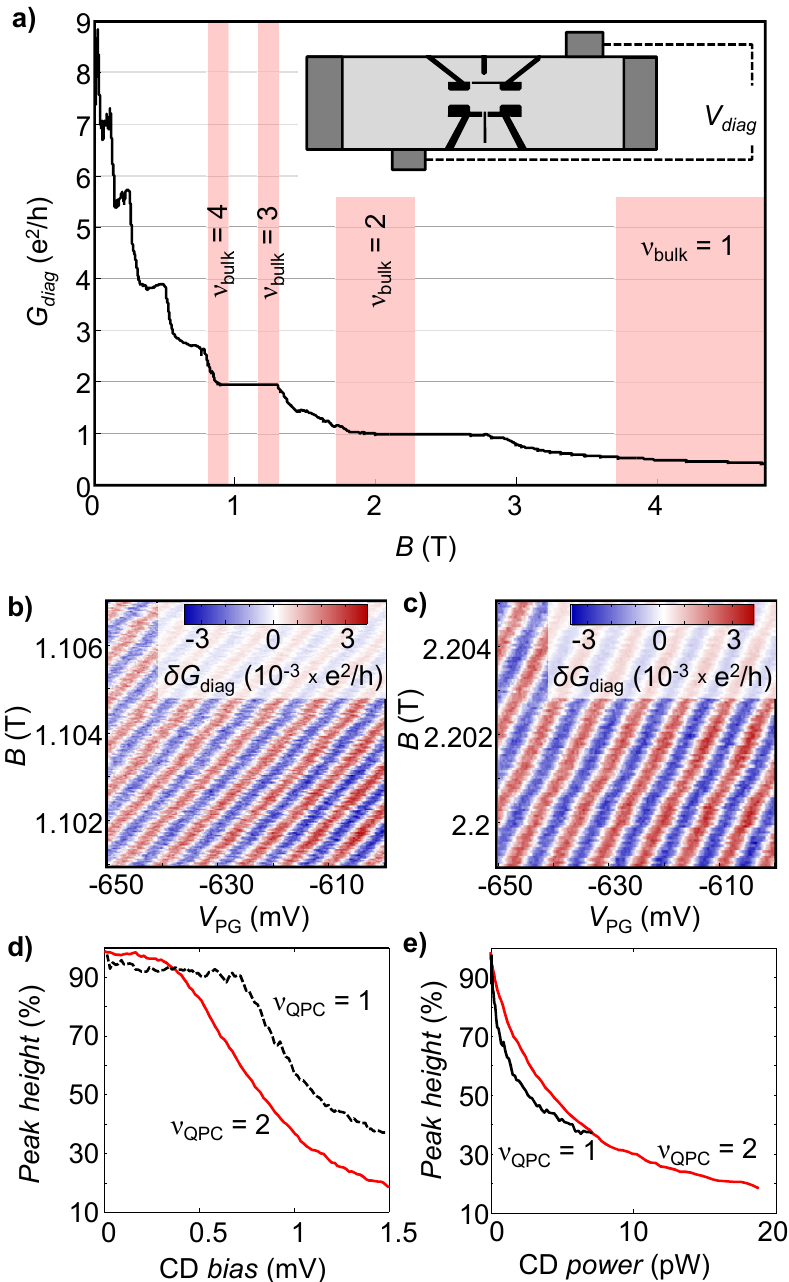}
\caption{\label{fig:4} \textbf{a):} From the Hall voltage drop $V_{diag}$ diagonally across the QD (inset), the effective conductance through the QD can be extracted. Here, we show $G_{diag}$ for a fixed voltage applied to the topgates, as a function of magnetic field. $G_{diag}$ is quantized in multiples of $\frac{e^2}{h}$. Shaded regions indicate the filling factor $\nu_{\mathrm{bulk}}$ of the bulk at the corresponding magnetic field values. On the riser of the conductance plateaus, magnetoresistance oscillations are observed. Their dependence on magnetic field and plunger gate voltages is shown in \textbf{b)} for $\nu_{\mathrm{QPC}}\approx 2$, $\nu_{\mathrm{bulk}}\approx 4$ and in \textbf{c)} for $\nu_{\mathrm{QPC}}\approx 1$, $\nu_{\mathrm{bulk}}\approx 2$. Increasing the bias and the power applied to the charge detector QPC respectively, greatly reduces the peak height of the observed magnetoresistance oscillations (\textbf{d),e)}).}
\end{figure}
The Hall voltage drop across the QD ($V_{\mathrm{diag}}$, see inset Fig. A1a) gives access to the conductance through both constrictions \cite{zhang_distinct_2009}. When both QPCs are tuned to the same transmission ($G\approx 10\frac{e^2}{h}$ at $B=0$) by applying a negative topgate voltage and the magnetic field is varied, we find conductances through the constrictions quantized in multiples of $\frac{e^2}{h}$ (Fig. A1a). In this configuration, edge states are formed in the bulk, in the QD and in the QPCs. In the QPCs, the filling factor is lower than in bulk and QD: $\nu_{\mathrm{QPC}}<\nu_{\mathrm{QD}},\nu_{\mathrm{bulk}}$ (for our large dot $\nu_{\mathrm{QD}}\approx\nu_{\mathrm{bulk}}$). We note that $\nu_{\mathrm{QPC}}$ edge states pass the QD and contribute to the diagonal conductance, while $(\nu_{\mathrm{QD}}-\nu_{\mathrm{QPC}})$ edge states are confined inside the QD.
On the riser of the conductance-plateaus (i.e. the low magnetic field side), periodic conductance oscillations are observed (Fig. A1b for $\nu_{\mathrm{QPC}}\approx 2$, Fig. A1c for $\nu_{\mathrm{QPC}}\approx 1$). The peaks of these oscillations are shifted to lower magnetic fields as the plunger gate voltage is decreased (Fig. A1b,c). The strength of this shift depends on the filling factor inside the QPCs ($\Delta B \approx 1.0$ mT for $\nu_{\mathrm{QPC}}\approx 2$, $\Delta B \approx 1.9$ mT for $\nu_{\mathrm{QPC}}\approx 1$ , a smooth background has been subtracted). However, the plunger gate spacing $\Delta V_{\mathrm{PG}}$ is similar for both cases ($\Delta V_{\mathrm{PG}}\approx$ 7.2 mV for $\nu_{\mathrm{QPC}}\approx 2$, $\Delta V_{\mathrm{PG}}\approx$ 7.9 mV for $\nu_\mathrm{{QPC}}\approx 1$).
This scaling with the QPC filling factor, as well as the direction of the shift are both contrary to what is expected for an Aharonov-Bohm interferometer \cite{halperin_theory_2011}. Instead, the results are consistent with previous experiments and show that transport is governed by a Coulomb blockade mechanism \cite{zhang_distinct_2009,camino_quantum_2007,ofek_role_2010}. In this picture, the slope of the magnetoconductance oscillations is caused by the capacitive coupling of confined and transmitted edge states in the QD instead of a direct effect of the plunger gate on the interferometer area as in the Aharonov-Bohm case. Increasing the bias applied to the charge detector QPC ($G\approx 0.2 \frac{e^2}{h}$, Fig. A1d) decreases the amplitude of the oscillations, while the background of the conductance approaches its plateau value.

\section*{References}
\bibliographystyle{iopart-num}
\bibliography{Bibliothek}

\end{document}